\documentclass[pdflatex,sn-mathphys-num]{sn-jnl}

\usepackage{graphicx}%
\usepackage{multirow}%
\usepackage{amsmath,amssymb,amsfonts}%
\usepackage{amsthm}%
\usepackage{mathrsfs}%
\usepackage[title]{appendix}%
\usepackage{xcolor}%
\usepackage{textcomp}%
\usepackage{manyfoot}%
\usepackage{booktabs}%
\usepackage{algorithm}%
\usepackage{algorithmicx}%
\usepackage{algpseudocode}%
\usepackage{listings}%

\theoremstyle{thmstyleone}%

\theoremstyle{thmstyletwo}%

\theoremstyle{thmstylethree}%

\begin{document}

\title[Calibration-Free Broadband Reconstructive Spectrometer]{Calibration-Free Broadband Reconstructive Spectrometer based on Conical Phase Modulation}

\author[1]{\fnm{Junrui} \sur{Liang}}

\author[1]{\fnm{Zhongming} \sur{Huang}}

\author[1]{\fnm{Yanting} \sur{Guo}}

\author[1,2,3]{\fnm{Jun} \sur{Ye}}

\author[1]{\fnm{Yanzhao} \sur{Ke}}

\author*[1]{\fnm{Jun} \sur{Li}}\email{lijun\_gfkd@nudt.edu.cn}

\author*[1]{\fnm{Jiangming} \sur{Xu}}\email{jmxu1988@nudt.edu.cn}

\author[1]{\fnm{Jinyong} \sur{Leng}}

\author*[1]{\fnm{Pu} \sur{Zhou}}\email{zhoupu@nudt.edu.cn}

\affil[1]{\orgdiv{College of Advanced Interdisciplinary Studies}, \orgname{National University of Defense Technology}, \orgaddress{\city{Changsha}, \postcode{410073}, \country{China}}}

\affil[2]{\orgdiv{Nanhu Laser Laboratory}, \orgname{National University of Defense Technology}, \orgaddress{\city{Changsha}, \postcode{410073}, \country{China}}}

\affil[3]{\orgdiv{Hunan Provincial Key Laboratory of High Energy Laser Technology}, \orgname{National University of Defense Technology}, \orgaddress{\city{Changsha}, \postcode{410073}, \country{China}}}

\abstract{Reconstructive spectrometers (RSs) shift spectral analysis from hardware-heavy dispersion to encoding--decoding schemes. In the speckle-based RSs, spectra are encoded into disordered intensity patterns and then recovered algorithmically. However, characterizing such spectral-encoding transmission matrices (TMs) typically demands empirically complex calibrations, which rely on expensive tunable light sources to generate known wavelengths. In this work, we demonstrate a fully calibration-free RS that leverages conical phase modulation from an axicon. The target spectra are encoded into a Bessel-form intensity profile, enabling computational construction of a deterministic TM and thus completely removing the need for optical calibration. With this computer-generated TM, broadband spectral reconstruction is experimentally achieved across a 650 nm range (1050-1700 nm)---nearly three times the bandwidth of existing calibration-simplified RSs---while a spectral resolution of 2.9 nm remains comparable to that of these counterparts. Compared with an RS requiring optical calibration for equivalent performance, our solution achieves an approximate 36-fold cost reduction. Unlike most speckle RSs, the axicon RS requires no fixed input polarizer, further enhancing energy efficiency and simplicity through polarization-insensitive operation. The calibration-free TM also unlocks a data-rich RS paradigm through low-cost computation. We show that neural networks trained purely on synthetic spectral-spatial data generalize directly to experiments, with rapid inference enabling 267-Hz near-infrared spectral classification. Benefiting from the dataset-level statistics, we further propose mutual-information-guided spatial sampling that enables accurate classification with as few as 4 detector pixels (2.45\% ultra-low sampling rate), yielding a relative 113\% accuracy gain over conventional uniform sampling. To the best of our knowledge, this is the first reported experimentally calibration-free RS, featuring superior simplicity, practicality, and cost-effectiveness for practical applications.}




\maketitle

\section{Introduction}\label{sec1}

\begin{figure}[t]
\centering
\includegraphics[width=\textwidth]{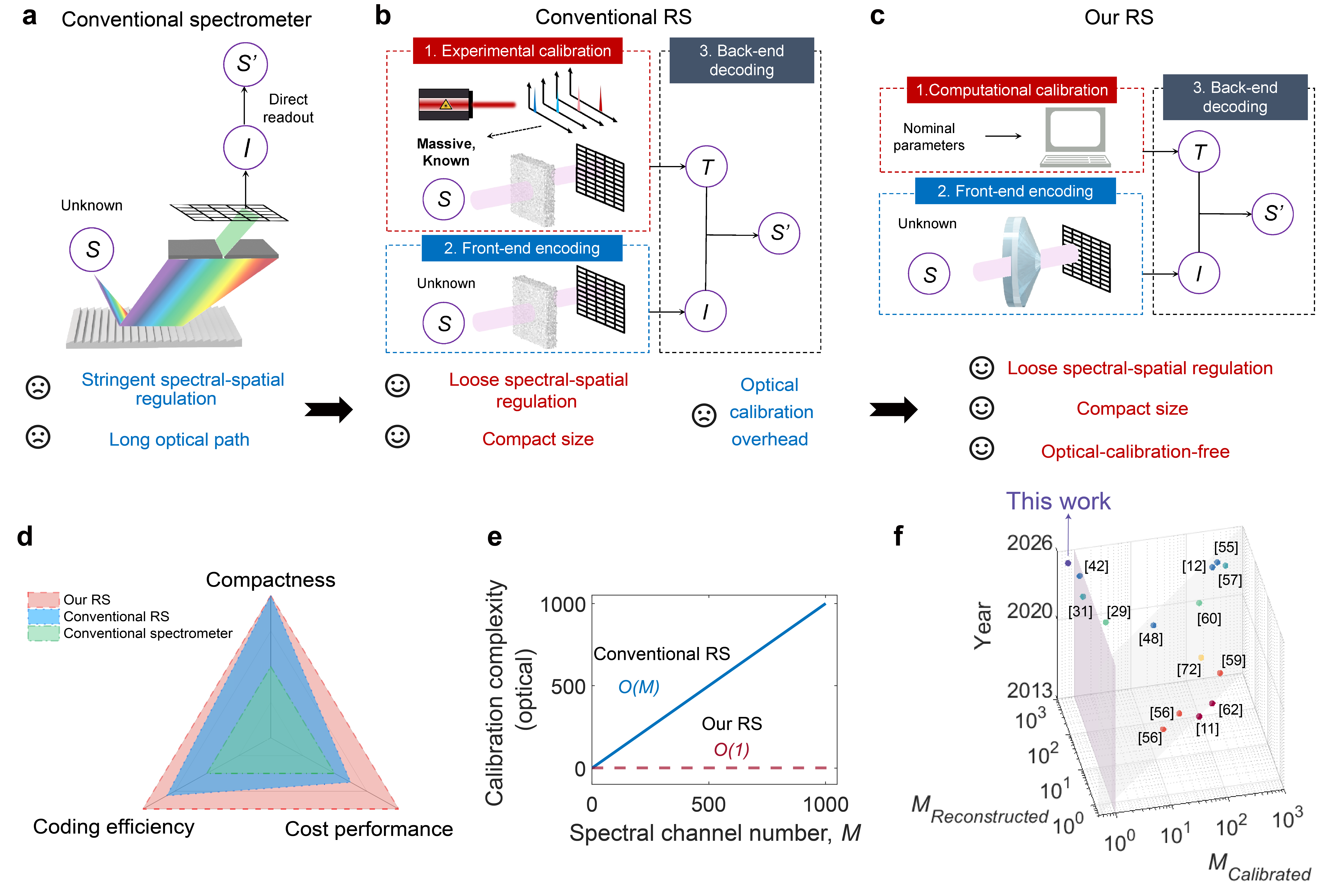}
\caption{Schematic of our proposed RS. \textbf{a}, In a conventional spectrometer, incident light is dispersed and propagates a long path before wavelengths are strictly concentrated onto corresponding spatial pixels, allowing the spectrum to be read out directly.  \textbf{b}, A conventional RS typically involves three steps: experimental calibration, front-end encoding, and back-end decoding. Before the unknown spectrum is entirely encoded into intensity signals by a compact device, a tunable light source generates massive known wavelengths to calibrate the encoder. An algorithm then reconstructs the spectrum from the calibration and measurement data.  \textbf{c}, Our RS employs an axicon for an analytically deterministic encoding, bypassing experimental calibration entirely. The encoding TM is readily derived from nominal device parameters via simple computation. \textbf{d}, Comparison of conventional spectrometers, conventional RS, and our RS in terms of compactness, coding efficiency, and cost-performance. \textbf{e}, Approximate relationship between optical calibration complexity and number of spectral channels. \textbf{f}, A brief survey of RS development, focusing on the relationship between the number of calibrated channels and reconstructed channels. The gray plane corresponds to the most cases with $M_{\text{Calibrated}}$  = $M_{\text{Reconstructed}}$, and the light purple plane to the existing cases with $M_{\text{Calibrated}}$ = 1. Dark purple dot represents this work with $M_{\text{Calibrated}}$ = 0. }
\label{fig:1}
\end{figure}

Spectrometers play an indispensable role in modern industry and scientific research, with widespread applications in material characterization, hyperspectral imaging, and other fields \cite{Yiru,Chunhui,3,4,5}. However, conventional methods are typically limited by large size, heavy weight, and high cost, which restrict their application in specific scenarios (see Fig.~\ref{fig:1}a). In recent years, advances in computational power and algorithms have enabled the development of reconstructive spectrometers (RSs) based on complex spectral-spatial mappings \cite{6,7,8,9,10}. This approach significantly reduces the burden of dispersive optics and allows for spectral analysis within a small footprint and lightweight structure. Generally, devices with distinctive wavelength-dependent responses can act as spectral encoders. Algorithms are capable of recovering or approximating the original spectrum from the observed signals, even if the encoded intensities are not strictly spatially separated and may even be mixed or scrambled. Disordered waveguides \cite{11,12}, fibers \cite{13,14}, metasurfaces \cite{15,16}, and chaotic optical cavities \cite{10,17,18} have been employed in such systems. Light propagating through these opaque media may undergo strong scattering \cite{19,20}, complex interference \cite{21,22}, energy coupling \cite{23,24}, or random diffraction \cite{15,16}, leading to an irregular speckle pattern at the detector---a spectral encoding process that is hard to characterize analytically. To make the back-end spectral decoding problem tractable, it is essential to empirically acquire the prior knowledge of the front-end encoding model (see Fig.~\ref{fig:1}b). This is typically achieved by calibrating the spectral-spatial transmission matrix (TM) using tunable light sources.

Specifically, TM calibration is required to span all target wavelength channels over the full operational range of the RS. Such a procedure is labor-intensive and resource-demanding, with the cost becoming even higher in special wavebands where tunable light sources are scarce \cite{25,26}. The calibration process also undermines the system's simplicity and practicality. Therefore, reducing or even removing the requirement for calibration has become a critical pursuit for RSs, and has attracted significant research attention in recent years. Previous work has attempted to achieve wavelength reconstruction at a finer resolution than the calibration resolution of existing devices \cite{18,27,28}. And another research trend is to explore how to minimize the number of calibration wavelengths. To enable spectral reconstruction with sparser calibration, a numerical model first optimized  from single-wavelength experimental calibration is utilized to predict speckle outputs of disordered metasurfaces across a 220-nm range \cite{29}. In contrast to disordered mappings, ordered one-to-many spectral-spatial mappings offer an alternative route. The parabolic mapping exhibited by a planar 1D photonic crystal cavity enables determining the entire TM using bichromatic calibration, with a limited working range to tens of nanometers \cite{29,30}. Advancing this direction, a pinhole-based RS scheme is proposed \cite{31}, which exploits the deterministic correspondence between the Fraunhofer diffraction pattern and the incident wavelength. A simple monochromatic calibration combined with an affine transformation allows constructing a TM over a 200 nm range. However, the amplitude restriction introduced by the micron-sized pinhole drastically degrades energy efficiency. Capturing the diffraction pattern even requires detection with ultra-high dynamic range. According to Fraunhofer diffraction theory \cite{32}, the far-field propagation of the light field follows a Fourier transform model, with signal intensity decaying rapidly from the center to the edges. The resulting dynamic range can reach a level of $10^7$ \cite{33}---beyond the capability of standard cameras.

Although the above efforts have enabled simplified calibration for RSs, achieving truly calibration-free operation remains a critical open challenge (see Fig.~\ref{fig:1}f). In this work, we demonstrate the first fully calibration-free RS system, which completely eliminates the need for experimental pre-calibration of the spectral encoder  (see Fig.~\ref{fig:1}c). By employing the conical phase modulation of an axicon, a deterministic and analytically tractable spectral-spatial mapping is established, thereby converting the incident spectrum into a Bessel-shaped intensity distribution. The spectral response function of the system can be directly derived from its nominal parameters, and its validity is thoroughly verified in our experiments. We realize broadband spectral reconstruction over a 650 nm bandwidth (1050--1700 nm), with a coefficient of determination ($R^2$) of 0.9999 relative to a commercial spectrometer. This operating bandwidth is nearly three times that of existing calibration-simplified RSs. Spectral lines separated by 2.9 nm can be resolved, a performance comparable to those systems. Compared with an RS requiring optical calibration for equivalent performance, our solution achieves an estimated cost reduction of about 36-fold (evaluated using the equipment in our laboratory). The proposed spectral reconstruction scheme uses commercially available components and features a compact monolithic architecture, thus avoiding sophisticated high-precision fabrication. In addition, the full-band encoding and phase-modulation mechanism, together with its polarization-insensitive response, ensures high optical throughput efficiency. The advantages and positioning of this work within the spectrometer community are summarized in Fig.~\ref{fig:1}d-f.

\section{Principles}\label{sec2}

\begin{figure}[htbp]
\centering
\includegraphics[width=0.85\textwidth]{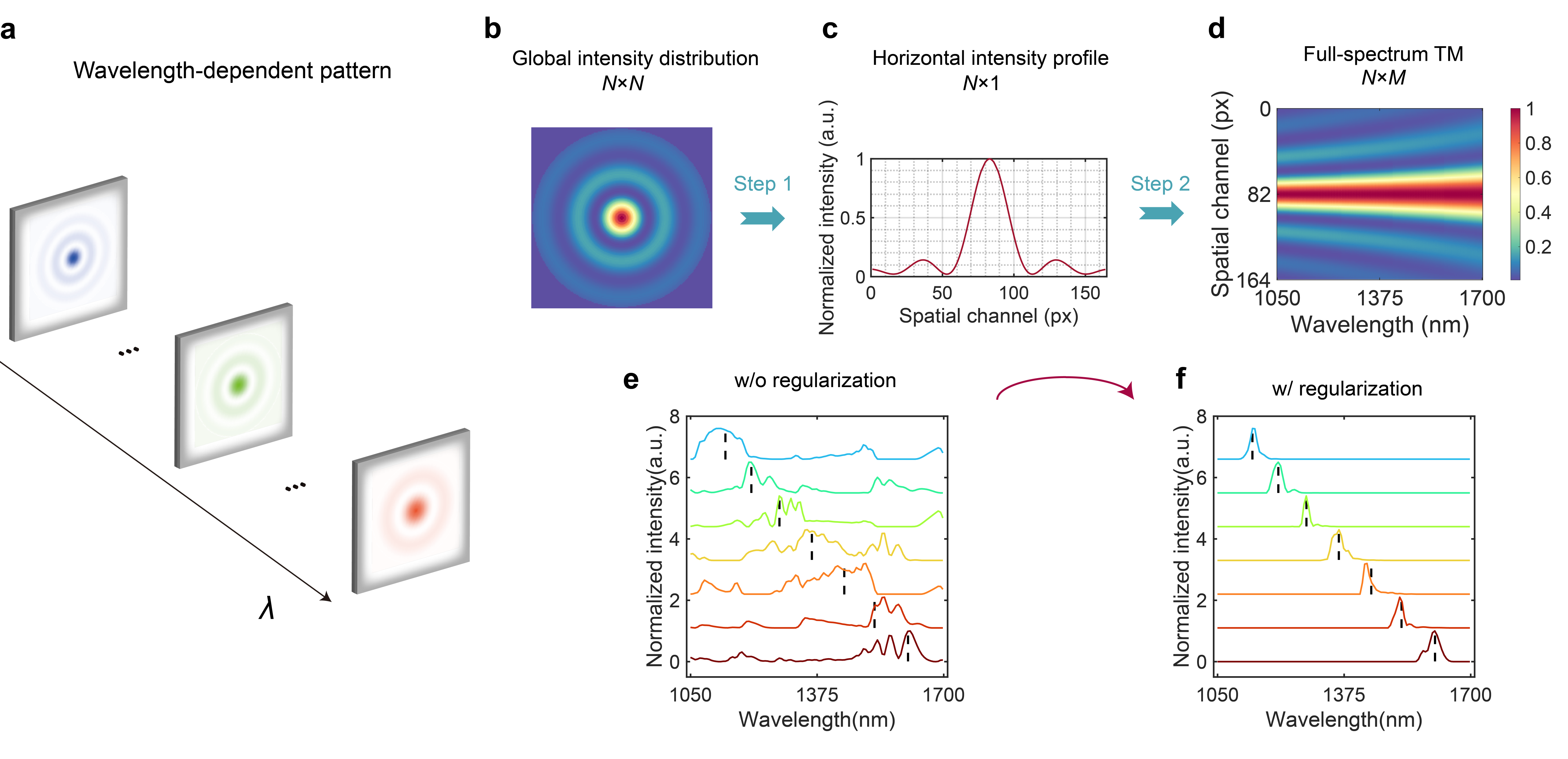}
\caption{Axicon for spectral encoding and synthetic spectrum reconstruction. \textbf{a}, The intensity distribution pattern after axicon modulation exhibits wavelength dependence. \textbf{b}, Simulated global intensity distribution corresponding to a single wavelength, with a size of $N \times N$. \textbf{c}, Simulated one-dimensional horizontal profile with a size of $N \times 1$. \textbf{d}, Full-spectrum TM generated by computer. Synthetic spectral reconstruction spanning 1050--1700~nm using \textbf{e}, the pseudoinverse, and \textbf{f}, the regularization-based algorithm.
}
\label{fig:2}
\end{figure}

The axicon is a widely used optical phase modulation device for applications including Bessel beam generation \cite{34}, optical coherence tomography \cite{35}, and so on. Considering the case of a parallel beam incident normally on an axicon, where the base angle of the axicon is $\alpha$, the maximum clear aperture radius is $R$, and the refractive index is $n$. A cylindrical coordinate system is established with the optical axis as the $z$-axis, and the polar coordinate plane passing through the apex of the axicon is defined as the $(r,\theta)$ plane. Under the paraxial approximation, the complex amplitude transmittance function of the axicon can be expressed as follows \cite{36,37,38}:
\begin{equation}
f(\lambda, r) =
\begin{cases}
\exp\left[-\frac{2\pi}{\lambda} j(n-1)\alpha r\right], & r \leq R \\
0, & r > R
\end{cases}
\end{equation}
where $j$ is the imaginary unit, $\lambda$ is the optical wavelength, and $r$ is the radial distance from the apex on the rear surface of the axicon. Furthermore, a transverse plane perpendicular to the optical axis behind the axicon is defined as the $(\rho,\phi)$ plane. By applying the Fresnel diffraction formula in cylindrical coordinates and under the simplification of a unit-amplitude, normally incident plane wave, the resulting intensity distribution at a longitudinal position $z$ behind the axicon is derived as (detailed derivation is provided in Supplementary S1):
where $J_0$ represents the zero-order Bessel function of the first kind. Based on Eq.~(2), we are able to simulate the wavelength-dependent spatial intensity pattern behind the axicon (see Method for details). As schematically illustrated in Fig.~\ref{fig:2}a, the diameter of the central spot and the distribution of concentric rings vary correspondingly with wavelength. Since the intensity distribution is independent of the azimuthal angle $\phi$ (as given by Eq.~(2)) and thus rotationally symmetric, we select only the horizontal radial cross-sectional profile for further analysis (Fig.~\ref{fig:2}b, c), which can greatly reduce data redundancy and allow for a higher acquisition frame rate \cite{39}. A full-band spectral-spatial TM can be further generated by arranging the 1-D intensity profiles column-wise (Fig.~\ref{fig:2}d).

The computer-generated TM is critical for spectral encoding and decoding process. For an unknown spectral input, the intensity distribution recorded by the sensor can be modeled as a weighted superposition of monochromatic patterns: $I(\rho) = \int T(\rho,\lambda) S(\lambda) d\lambda$, where $I(\rho)$ represents the intensity at position $\rho$ of the imaging sensor, $S(\lambda)$ denotes the power at wavelength $\lambda$, and $T(\rho,\lambda)$ is the device's spectral response at $\rho$ for wavelength $\lambda$. In discretized form, this relationship simplifies to a linear system:
\begin{equation}
\mathbf{I} = \mathbf{T}\mathbf{S}
\end{equation}
where $\mathbf{I} \in \mathbb{R}^{N\times 1}$ is a vector representing measured intensity at each pixel of the sensor and $\mathbf{S} \in \mathbb{R}^{M\times 1}$ is a vector representing the discretized spectrum. $N$ and $M$ denote the observed spatial channel number and the spectral channel number under test. $\mathbf{T} \in \mathbb{R}^{N\times M}$ is the computer-generated TM.

In a conventional spectrometer, the \textbf{T} is an identity matrix, and each wavelength contributes intensity only at its specific spatial location; namely, the mapping is one-to-one.  In contrast, in this work, \textbf{T} distributively contains a large number of nonzero entries, such that each wavelength component contributes intensity at all spatial positions; hence, there is a one-to-many mapping. Moreover, the raw measurements here carry no direct physical meaning, and the probe spectrum must be reconstructed algorithmically. Following the approach in the studies \cite{10,11} , we first synthesize the observed intensities using simulated spectra and a TM perturbed by added Gaussian noise, thereby validating the effectiveness of the spectral reconstruction algorithm in our axicon encoding system. According to Eq.~(3), the unknown spectrum $\mathbf{S}$ can be recovered by multiplying the observed $\mathbf{I}$ by the inverse of the known $\mathbf{T}$: $\mathbf{S} = \mathbf{T}^{+} \mathbf{I}$. Here, $^{+}$ denotes the Moore-Penrose pseudoinverse, as $\mathbf{T}$ is not necessarily square. In practice, the ill-posedness of the equation system renders the matrix inversion susceptible to noise, leading to inaccurate results (see Fig.~\ref{fig:2}e for reconstructing widely tunable spectra).  To improve the reconstruction accuracy, we introduce an additional regularizer and solve the following optimization problem:
\begin{equation}
\min_{\mathbf{S} \ge 0} \left\{ \|\mathbf{I} - \mathbf{T}\mathbf{S}\|_2^2 + \Gamma(\mathbf{S}) \right\}
\end{equation}
Here, the non-negativity constraint ensures physical plausibility of the recovered spectral values.  $\Gamma(\mathbf{S})$ represents the regularization term. As an illustration, by applying $\ell_1$-norm regularization to enforce sparsity in a broadband range, the spectral reconstruction performance is improved from Fig.~\ref{fig:2}e  to Fig.~\ref{fig:2}f  under the same measurement conditions.  Based on the priors of spectra in terms of smoothness and gradient, $\ell_2$-norm regularization and total variation regularization can also be employed \cite{Yiru}. For detailed implementation of our regularization algorithms, please refer to the Methods section and Supplementary Information S2.

\section{Results}\label{sec3}

\begin{figure}[htbp]
\centering
\includegraphics[width=0.85\textwidth]{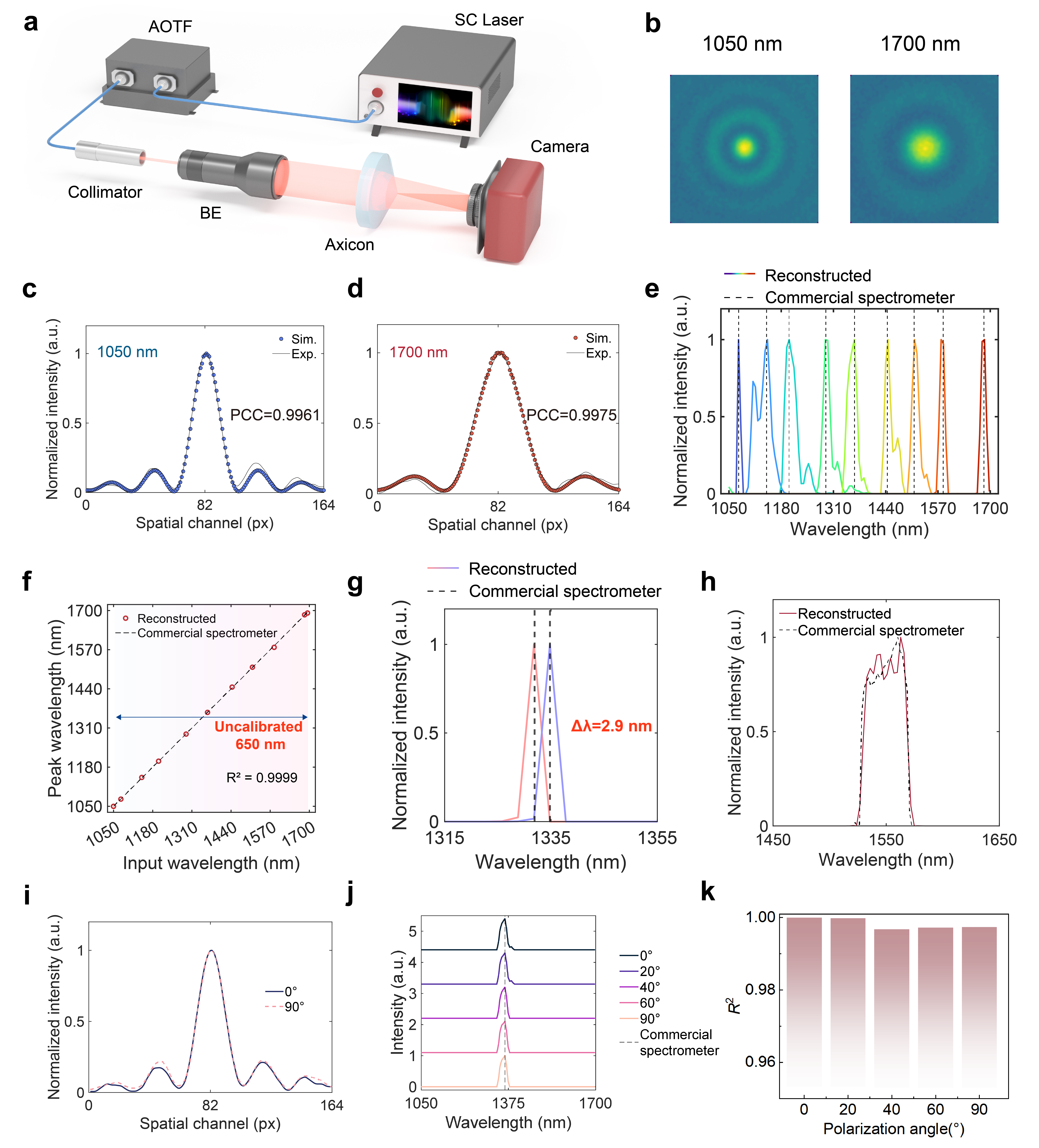}
\caption{The axicon-based RS is applied for spectral measurements. \textbf{a}, Experimental setup of the axicon-based RS. SC: supercontinuum; AOTF: acousto-optic tunable filter; BE: beam expander. \textbf{b}, Observed intensity pattern corresponding to 1050 nm and 1700 nm. Simulated and observed 1D intensity profiles corresponding to \textbf{c}, 1050 nm and \textbf{d}, 1700 nm. \textbf{e}, Broadband spectral reconstruction spanning 1050--1700 nm. \textbf{f}, Comparison between the recovered peak wavelengths and the input peak wavelengths measured by a commercial spectrometer. \textbf{g}, Reconstructed spectra with a step size of 2.9 nm. \textbf{h}, Reconstructed spectrum of an ASE light source. \textbf{i}, Measured intensity distributions after axicon modulation under 0$^\circ$ and 90$^\circ$ incident polarization states. \textbf{j}, Spectral reconstruction results under different incident polarization angles. \textbf{k}, The $R^2$ for peak wavelength reconstruction in broadband spectral measurement as a function of the incident polarization angle.}
\label{fig:3}
\end{figure}

In this section, we verify the capability of using purely computer-generated TM for spectral reconstruction in experiments, with the setup presented in Fig.~\ref{fig:3}a (see Method for details). Figure~\ref{fig:3}b displays the 2D observed intensity pattern corresponding to 1050 nm and 1700 nm. Figures~\ref{fig:3}c,d show a comparison between the simulated and experimentally observed horizontal intensity profiles at these two representative wavelengths, with Pearson correlation coefficients (see Method for definition) as high as 0.9961 and 0.9975, respectively.

For the purpose of comparison, the produced spectra are incident onto both our RS and a commercial spectrometer, respectively. First, we demonstrate the wide operating bandwidth of the system. As illustrated in Fig.~\ref{fig:3}e, effective reconstruction is realized across the range of 1050--1700 nm. While the slower spectral variation of the concentric intensity rings at shorter wavelengths (see Fig.~\ref{fig:2}d) broadens the reconstructed peaks, it still allows for sufficiently indicative results across the wide range. We employ the coefficient of determination $R^2$ (see definition in Method), a widely adopted metric in the field of RS \cite{10,13,40}, to evaluate the agreement between the reconstructed central wavelength values and the wavelength values under test. The $R^2$ can be as high as 0.9999, as depicted in Fig.~\ref{fig:3}f. The current operational bandwidth is primarily limited by the spectral response range of the detector array and the tunable wavelengths in our laboratory. Then, following the reported methods \cite{17,41}, we evaluate the spectral resolution by sequentially injecting single-peak spectra with a tiny separation into our RS. As shown in Fig.~\ref{fig:3}g, spectral peaks separated by 2.9 nm can be clearly distinguished, with the first zero point of one spectral line coinciding with the main maximum of the other, and their respective central positions are consistent with the measurements from a commercial spectrometer. This resolution is comparable to the state-of-the-art calibration-simplified RSs \cite{31,42}, and could be improved in the future by employing higher resolution detector arrays. Moreover, the spectrum of an amplified spontaneous emission (ASE) source with an FWHM of 41 nm is reconstructed, and the resulting spectral shape closely matches the measurements obtained using a commercial spectrometer (see Fig.~\ref{fig:3}h).

The axicon-based RS offers the advantage of low polarization sensitivity. In many other types of RSs, the core dispersive medium exhibits polarization-dependent response \cite{9,12,19}. To ensure that the system measures only a single variable, i.e., the wavelength, it is necessary to control the incident light to a single linear polarization state \cite{7,8}. This constraint not only increases hardware complexity but also blocks a portion of the incident optical energy. The measured intensity profiles in Fig.~\ref{fig:3}i demonstrate that the axicon is insensitive to the polarization state of the incident light. Spectral reconstruction results under different polarization angles show strong robustness, as illustrated in Fig.~\ref{fig:3}j. Figure~\ref{fig:3}k shows the variation of the $R^2$ for spectral reconstruction with respect to polarization angle in broadband spectral measurements, indicating that reconstruction consistently maintains high accuracy ($R^2>0.9960$).

\begin{figure}[htbp]
\centering
\includegraphics[width=\textwidth]{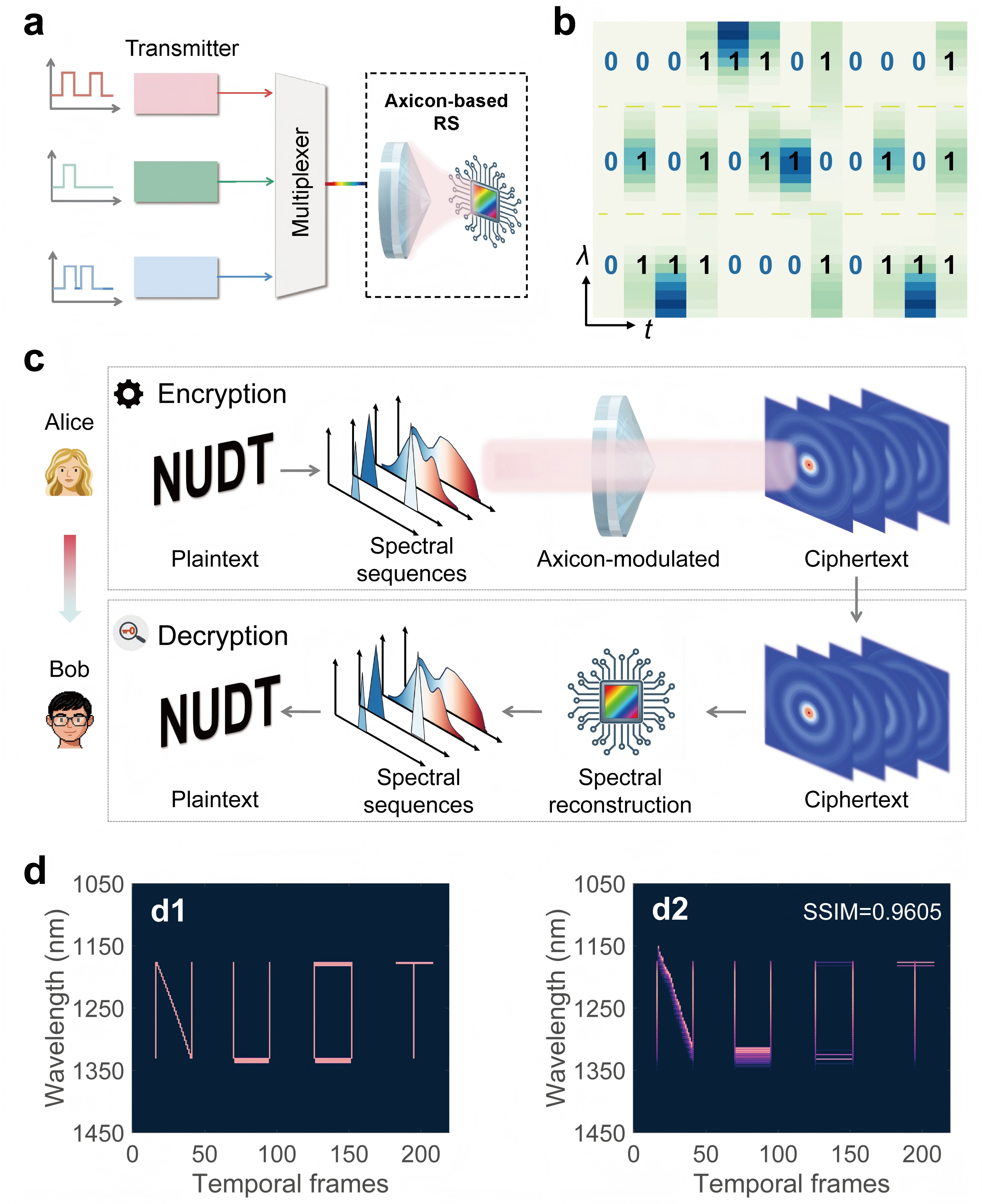}
\caption{Application of the proposed RS in optical communication and encryption. \textbf{a}, Diagram of the WDM optical communication system incorporating the axicon-based RS. \textbf{b}, The axicon-based RS demultiplexes the three-waveband WDM signals and decodes the original binary data streams. \textbf{c}, Schematic illustration of the axicon-based RS for optical encryption. \textbf{d}, Transmitted spectral pattern and its corresponding decryption result.}
\label{fig:4}
\end{figure}

We further demonstrate the practical application potential of the axicon-based RS in wavelength division multiplexing (WDM) communication. As schematized in Fig.~\ref{fig:4}a, we employ multiple independently modulated light sources at discrete bands near 1 $\mu$m, 1.3 $\mu$m, and 1.6 $\mu$m as transmitters. These signals are coupled into our axicon-based system, and the modulated spectral distribution is retrieved. The proposed RS clearly distinguishes the original binary modulation information (see Fig.~\ref{fig:4}b), and its broad operating range may enable application across various data transmission media \cite{43,44}.

Moreover, the relationship between the observed intensity distribution and the incident spectra in the RS is non-intuitive. This property opens up the possibility of using our system for optical encryption. As shown in Fig.~\ref{fig:4}c, at the encryption end, the plaintext (the pattern ``NUDT'') is first projected into a spectral sequence. The sender (Alice) then sequentially generates these spectral slices and illuminates the axicon (see Method for details), which imposes phase modulation to produce the ciphertext (spectrum-dependent intensity patterns) on the sensor array. After recording the ciphertext and being authorized by Alice to acquire the spectral-spatial TM as well as the reconstruction algorithm, the receiver (Bob) can recover the time-evolving spectrum. Connecting these spectral slices decrypts the plaintext (the pattern ``NUDT''). Fig.~\ref{fig:4}d1 shows the encoded time-varying spectral sequence at the transmitter, and Fig.~\ref{fig:4}d2 presents the reconstructed spectrum at the decryption end. The structural similarity (SSIM) index is employed to evaluate the decryption results (see definition in Method). The SSIM between the decrypted and plaintext patterns attains a value of 0.9605, indicating high decryption quality.

We have demonstrated that, based on the computationally generated TM, spectra can be reconstructed by iteratively solving a system of linear equations. However, the equation-solving approach has certain limitations, such as insufficient real-time capacity and degraded accuracy under low spatial sampling rates where the system of equations is severely underdetermined \cite{59}. In applications such as broadband laser warning \cite{45,article}, it is desirable to rapidly acquire indicative, though not necessarily highly precise, spectral information. Moreover, to facilitate flexible deployment of devices, it is essential to retrieve spectral information using a small number of detector units.

To this end, a neural network is employed in this study for the spectral decoding of the axicon RS. Unlike existing spectral decoding networks, which require experimentally acquired intensity samples and ground-truth spectral labels to provide training datasets \cite{4,50,56,3}, we can conveniently produce numerous training data that closely matches the realistic system using the computer-generated TM. Specifically, a series of synthetic spectral labels $\lambda$ are first generated and then mapped through the TM to obtain the corresponding $\mathbf{I}$. A tiny fully-connected neural network is adopted for this task (architecture and training details are provided in Methods). The network trained on synthetic data is then tested on real-world data.

We first test wide-spectrum classification with the infrared camera operating at 33~frames per second (FPS), achieving an accuracy as high as 100\%. When the detector frame rate is increased to 267~FPS, the classification accuracy drops to 92.5\% due to the reduced signal-to-noise ratio (SNR) of the captured intensity distribution caused by shorter exposure time. Given that the single inference time of the neural network (approximately 0.1~ms) is far shorter than the exposure time per frame (3.75~ms), the spectral information can practically be acquired at a rate of 267~Hz.

The above tests are conducted using the complete horizontal intensity profile (i.e., $N=164$ pixels), within which some pixels may carry redundant information. Existing RS studies have employed heuristic methods, including uniform sampling \cite{Redding:12} and local sampling \cite{54}, to find compressed subspaces of the raw measurements. The field still lacks a principled method for determining the most characterizing spatial features in spectral extraction.

\backmatter

\bmhead{Supplementary information}

Supplementary information accompanies this paper.

\bmhead{Acknowledgements}

The authors thank Mingkang Yuan and Haoyu Liu for valuable discussions and help on the video production.

\section*{Declarations}

\subsection*{Funding}
Innovation Research Foundation of National University of Defense Technology.

\subsection*{Conflict of interest}
The authors declare no conflicts of interest.

\subsection*{Data availability}
Data underlying the results presented in this paper are not publicly available at this time but may be obtained from the authors upon reasonable request.

\subsection*{Code availability}
Not applicable.

\subsection*{Author contributions}
J.L. conceived the idea and designed the experiments. Z.H. and Y.G. performed the numerical simulations. J. Ye and Y.K. contributed to the experimental setup. J. Li, J. Xu, J. Leng, and P.Z. supervised the project. All authors discussed the results and contributed to the manuscript.

\bibliography{ref}

\end{document}